\documentclass[preprint]{aastex}
\usepackage{multirow}

\newcounter{ionctr}
\ifx \undefined \ion \newcommand{\ion}[2]{\setcounter{ionctr}{#2}{#1$\;${\small\rmfamily\Roman{ionctr}}\relax}} \fi

\newcommand{\CID}{C(\ensuremath{^1}D)}

\newcommand{\water}{H\ensuremath{_2}O}

\ifx \undefined \lya \newcommand{\lya}{Ly\ensuremath{\alpha}} \fi

\ifx \undefined \arcsec \newcommand{\arcsec}{\mbox{$^{\prime\prime}$}} \fi%
\ifx \undefined \arcmin \newcommand{\arcmin}{\mbox{$^{\prime}$}} \fi%
\ifx \undefined \arcdeg \newcommand{\arcdeg}{\mbox{$^{\circ}$}} \fi%
\ifx \undefined \kms \newcommand{\kms}
	{\ifmmode{\mathrm{km\,s}^{-1}}\else{km\,s$^{-1}$}\fi}
	\fi%
\ifx \undefined \sysIII \newcommand{\sysIII}{\ensuremath{\lambda_\mathrm{III}}} \fi%
\ifx \undefined \sysIV \newcommand{\sysIV}{\ensuremath{\lambda_\mathrm{IV}}} \fi%
\ifx \undefined \quarterkev \newcommand{\quarterkev}{{${{1}\over{4}}$\,keV}}\fi%
\ifx \undefined \threequarterkev \newcommand{\threequarterkev}{{${{3}\over{4}}$\,keV}}\fi%

\newcommand{\RIo}{\mbox{\ifmmode{\,{$\mathrm{R_{Io}}$}}\else{\thinspace{R$_\mathrm{Io}$}}\fi}}%
\newcommand{\Rj}{\mbox{\ifmmode{\,{$\mathrm{R_{j}}$}}\else{\thinspace{R$_\mathrm{j}$}}\fi}}%
\newcommand{\by}{\mbox{$\times$}}%
\newcommand{\cmmtwo}{\ifmmode{\,{\rm cm}^{-2}}\else{\thinspace cm$^{-2}$}\fi}

\newcommand{\fig}[1]{Figure~\ref{fig:#1}}
\newcommand{\figp}[1]{Fig.~\ref{fig:#1}}

\newcommand{\figsp}[2]{Figs.~\ref{fig:#1}--\ref{fig:#2}}
\newcommand{\tab}[1]{Table~\ref{tab:#1}}

\newcommand{\about}{\ifmmode{\sim}\else{$\sim$}\fi}

\newcommand{\GALEX}{\textsl{GALEX}}
\newcommand{\IUE}{\textsl{IUE}}

\newcommand{\MSX}{\textsl{MSX}}
\newcommand{\WISP}{\textsl{WISP}}
\newcommand{\HST}{\textsl{HST}}

\newcommand{\Ulysses}{\textsl{Ulysses}}

\newcommand{\SOLSTICE}{\textsl{SOLSTICE}}
\newcommand{\Polar}{\textsl{Polar}}
\newcommand{\SOHO}{\textsl{SOHO}}

\shorttitle{The Ionization Lifetime of Carbon}
\shortauthors{Morgenthaler et al.}

\begin{document}

\title{\GALEX\ FUV Observations of Comet C/2004 Q2 (Machholz): The
Ionization Lifetime of Carbon}

\slugcomment{{\sc Accepted to ApJ:} October 23, 2010}

%

\author{Jeffrey P. Morgenthaler}
\affil{Planetary Science Institute, 1700 E.\ Fort Lowell, Ste 106,
  Tucson, AZ 85719, USA}
\email{jpmorgen@psi.edu}
\author{Walter M. Harris}
\affil{Department of Applied Sciences, University of California at
  Davis, One Shields Ave., Davis, CA 95616, USA}
\author{Michael R. Combi}
\affil{Department of Atmospheric, Oceanic and Space Sciences, The
University of Michigan, 2455 Hayward St.  Ann Arbor, MI 48109, Ann
Arbor, MI, USA}

\author{Paul D. Feldman}
\affil{Department of Physics and Astronomy, The Johns Hopkins University\\ 
Charles and 34th Streets, Baltimore, Maryland 21218, USA}

\author{Harold A. Weaver}
\affil{Space Department,
Johns Hopkins University Applied Physics \mbox{Laboratory,}
11100 Johns Hopkins Road,
Laurel, MD 20723-6099, USA}

\begin{abstract}
We present a measurement of the lifetime of ground state atomic
carbon, C($^3$P), against ionization processes in interplanetary space
and compare it to the lifetime expected from the dominant physical
processes likely to occur in this medium.  Our measurement is based on
analysis of a far ultraviolet (FUV) image of comet C/2004 Q2
(Machholz) recorded by the \textit{Galaxy Evolution Explorer} (\GALEX)
on 2005 March~1.  The bright \ion{C}{1} 1561\,\AA\ and 1657\,\AA\
multiplets dominate the \GALEX\ FUV band.  We used the image to create
high signal-to-noise ratio radial profiles that extended beyond
1$\times$10$^6$\,km from the comet nucleus.  Our measurements yielded
a total carbon lifetime of 7.1 -- 9.6$\times$10$^5$\,s (ionization
rate of 1.0 -- 1.4$\times$10$^{-6}$\,s$^{-1}$) when scaled to 1\,AU.
This compares favorably to calculations assuming solar
photoionization, solar wind proton change exchange and solar wind
electron impact ionization are the dominant processes occurring in
this medium and that comet Machholz was embedded in the slow solar
wind.  The shape of the \ion{C}{1} profiles inside 3$\times$10$^5$\,km
suggests that either the CO lifetime is shorter than previously
thought and/or a shorter-lived carbon-bearing parent molecule, such as
CH$_4$ is providing the majority of the carbon in this region of the
coma of comet Machholz.

\end{abstract}


\section{Introduction}
\label{intro}


The volatile content of comets makes their compositional study via
remote sensing particularly interesting and effective.  Dozens of
molecules and their elemental constituents have been detected in
cometary coma and ion tail emissions \citep[see
e.g.,][]{bockelee-morvan04, feldman04}.  Determining cometary nuclear
composition from these remote sensing studies requires detailed
understanding of the physical processes that govern coma and ion tail
emissions.  The study we report on here concerns one of these physical
processes: the ionization of ground state atomic carbon, C($^3$P).

UV emission from atomic carbon in comets was first detected in
sounding rocket observations of comet C/1973 E1
\citep[Kohoutek;][]{feldman74, opal74}.  The emission consists of
resonant fluorescence from C($^3$P), with multiplets at 1561\,\AA\ and
1657\,\AA\ (see e.g., Figure~2 of \citealt{festou84} or Figure~1 of
\citealt{tozzi98}).  These emissions are extremely extended spatially
\citep[e.g., Figure~1 of ][]{mcphate99}, implying a very long lifetime
of \ion{C}{1} against ionization processes.  The distribution of the
\ion{C}{1} emission rules out a direct nuclear source.  For most
comets, CO and CO$_2$ carry the bulk of carbon into cometary com\ae\
\citep[e.g., Table~1 of ][]{bockelee-morvan04}.  CH$_4$, C$_2$H$_2$,
C$_2$H$_6$, CH$_3$OH, and H$_2$CO also contribute.  As we discuss in
detail in \S\ref{analysis}, the final radial velocity of a carbon atom
liberated from any of these molecules is \about 4\,\kms.  The
lifetimes of the parent molecules are long enough to place the
liberated carbon well beyond the collision sphere of any comet
observed in the modern era.  Thus, the average carbon atom released
from a typical comet becomes an excellent test particle to probe
conditions in interplanetary space unperturbed by the presence of the
comet or any other solar system body.  By observing the radial
profiles of \ion{C}{1} in comets and fitting them with coma models, we
measure the lifetime of atomic carbon against ionization processes in
interplanetary space.

The majority of the \ion{C}{1} measurements to date have been
conducted by instruments on sounding rockets, \IUE\ and \HST, which
have small fields of view (FOVs) relative to the typical angular
extent of the \ion{C}{1} emission.  Exceptions to this have been
objective grating spectra of comet C/1975 V1-A
\citep[West][]{smith80}, narrowband wide-field images of comet C/1995
O1 (Hale-Bopp) recorded by \Polar\ spacecraft \citep{brittnacher01},
Wide Field Imaging Survey Polarimeter (\WISP) sounding rocket
observations of comet Hale-Bopp \citep{harris99}, and
imaging/spectro-imaging observations of Hale-Bopp by \MSX\
\citep{kupperman99T}.  With a technique similar to the one we use
here, \citeauthor{kupperman99T} found that the lifetime of atomic
carbon was consistent with the solar photoionization lifetime of
carbon calculated by \citet{huebner92}.  However, comet Hale-Bopp was
a comet of unusual size.  Several physical processes, including
collisional heating in the dense inner coma \citep{combi99, harris02}
and gas phase chemical interactions between coma species
\citep[e.g.,][]{glinski04} acted to extend the distribution of coma
species beyond what was expected using coma models based on comets
with more typical production rates.  \citeauthor{kupperman99T} did not
consider these effects.  We use observations of comet C/2004 Q2
(Machholz), which is a much more ``typical'' comet.  These
observations suggest that the lifetime of carbon against ionization is
about half of the photoionization lifetime calculated by
\citet{huebner92}.  This implies that another process, which we show
to be solar wind ionization, is comparable.


\section{Observations}
\label{obs}
The Galaxy Evolution Explorer (\GALEX) is a NASA Small Explorer (SMEX)
mission designed to map the history of star formation in the universe
\citep{martin05, morrissey05, morrissey07}.  Its 1.2\arcdeg\ FOV
diameter and high sensitivity to extended sources also makes \GALEX\
well suited to cometary coma studies.  \GALEX\ operates in two modes:
direct imaging and objective grism imaging.  Using a beam splitter and
two detectors, \GALEX\ images are simultaneously recorded in two
bands: 1350--1750\,\AA\ (FUV) and 1750--3100\,\AA\ (NUV).

\GALEX\ made both direct-mode and grism-mode observations of comet
C/2004 Q2 (Machholz).  The comet parameters are shown in \tab{obs}.
Over the course of the observations, the comet moved a substantial
fraction of the \GALEX\ 1.2\arcdeg\ FOV, so we quote only the center
of the spacecraft FOV, which was used for all observations.  A log of
observations is given in \tab{log}, which includes a background
direct-mode image that was recorded a month earlier.  Pipeline
processing did not complete properly for grism orbit~3 due to a
telemetry dropout, so we excluded it from our analyses.

\begin{figure}
  \plottwo{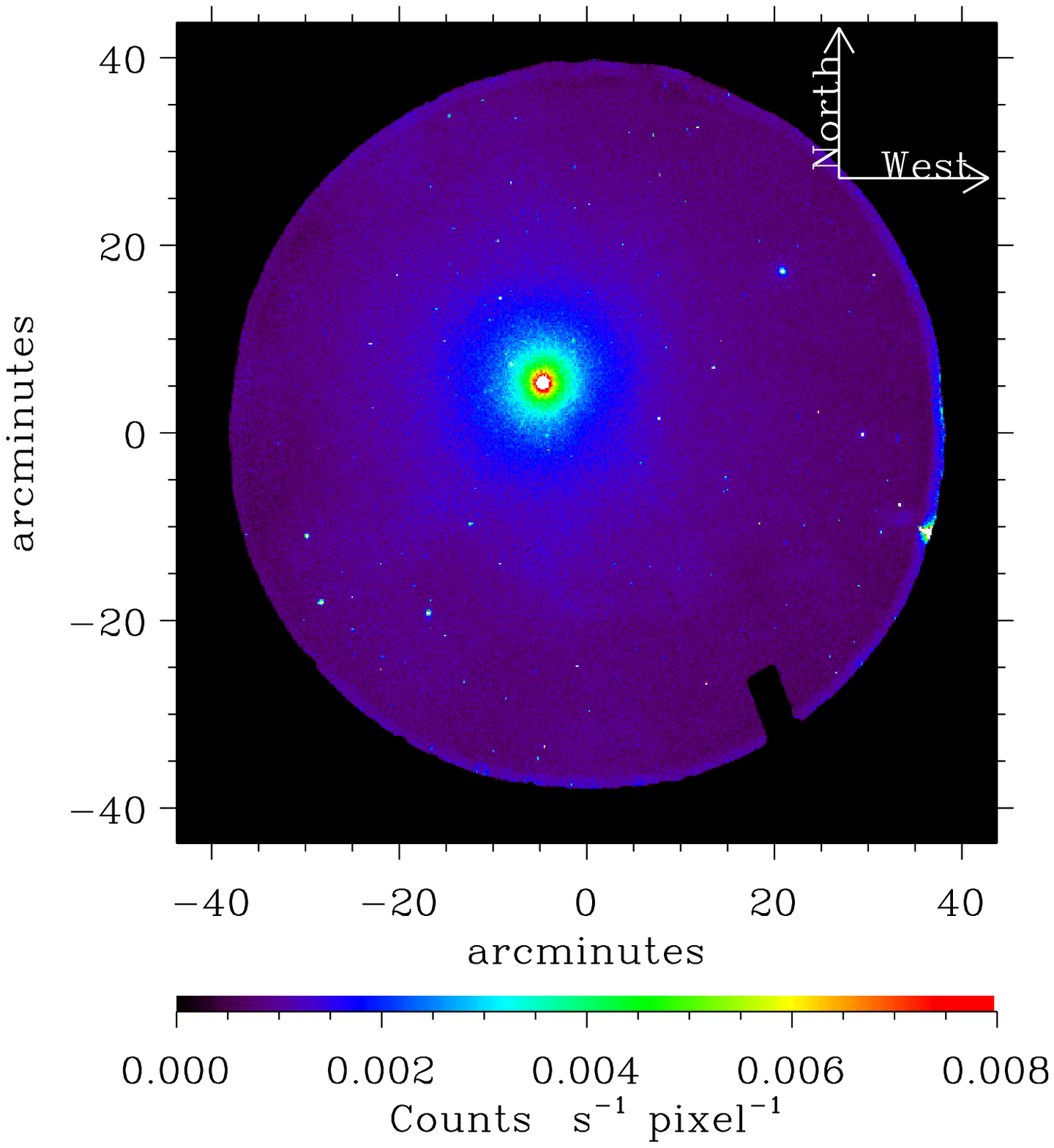}{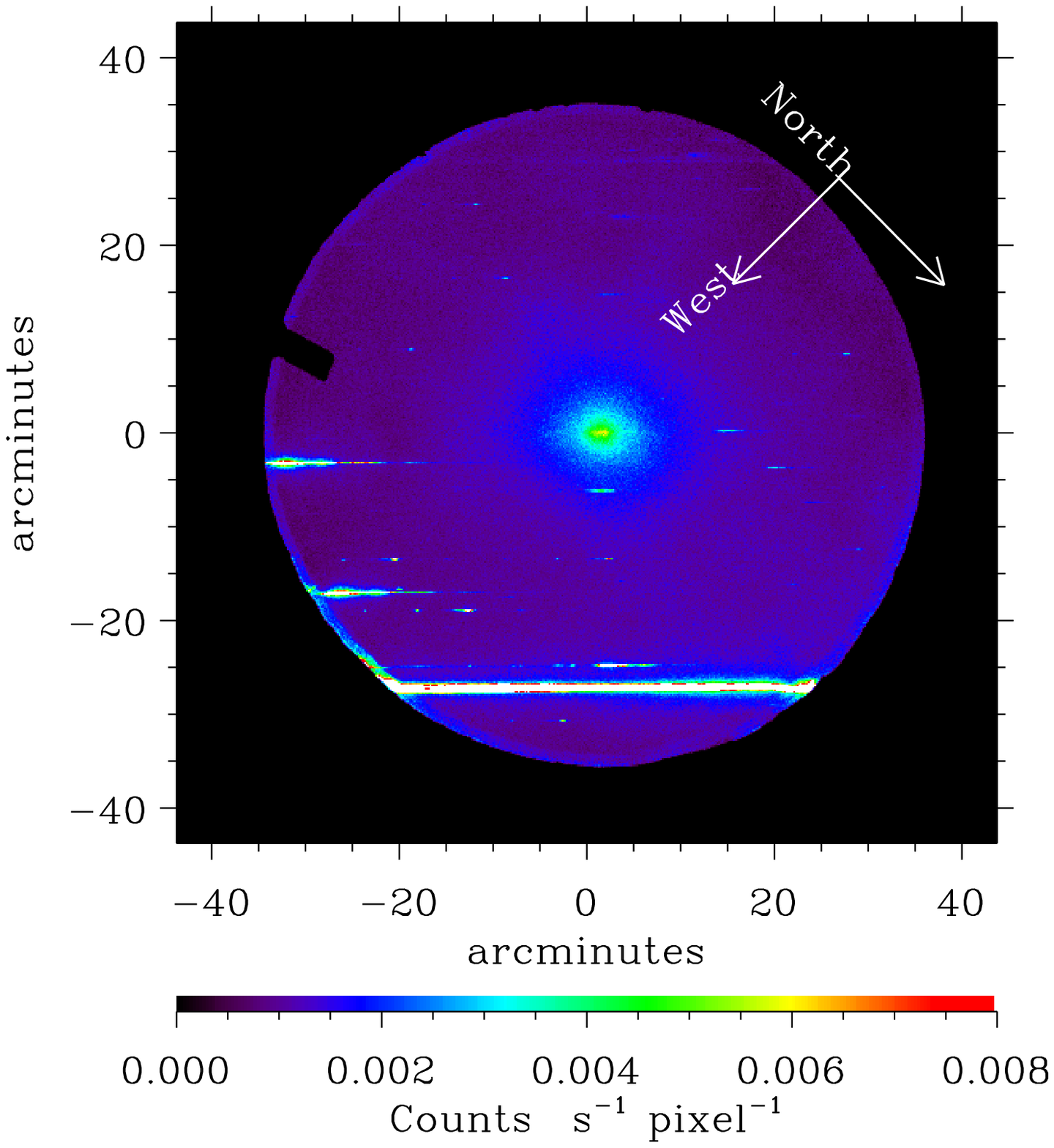}\\
  \plottwo{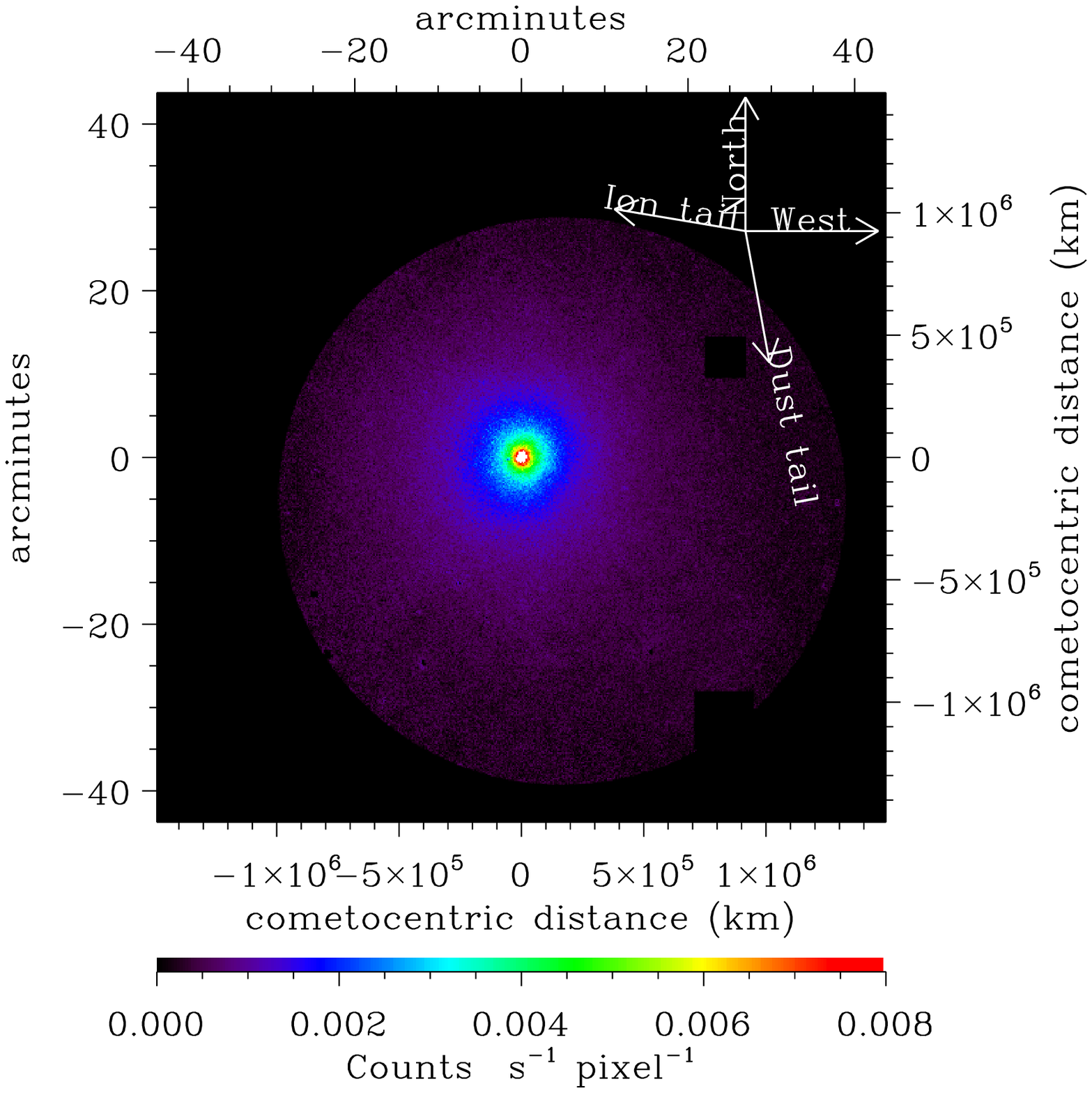}{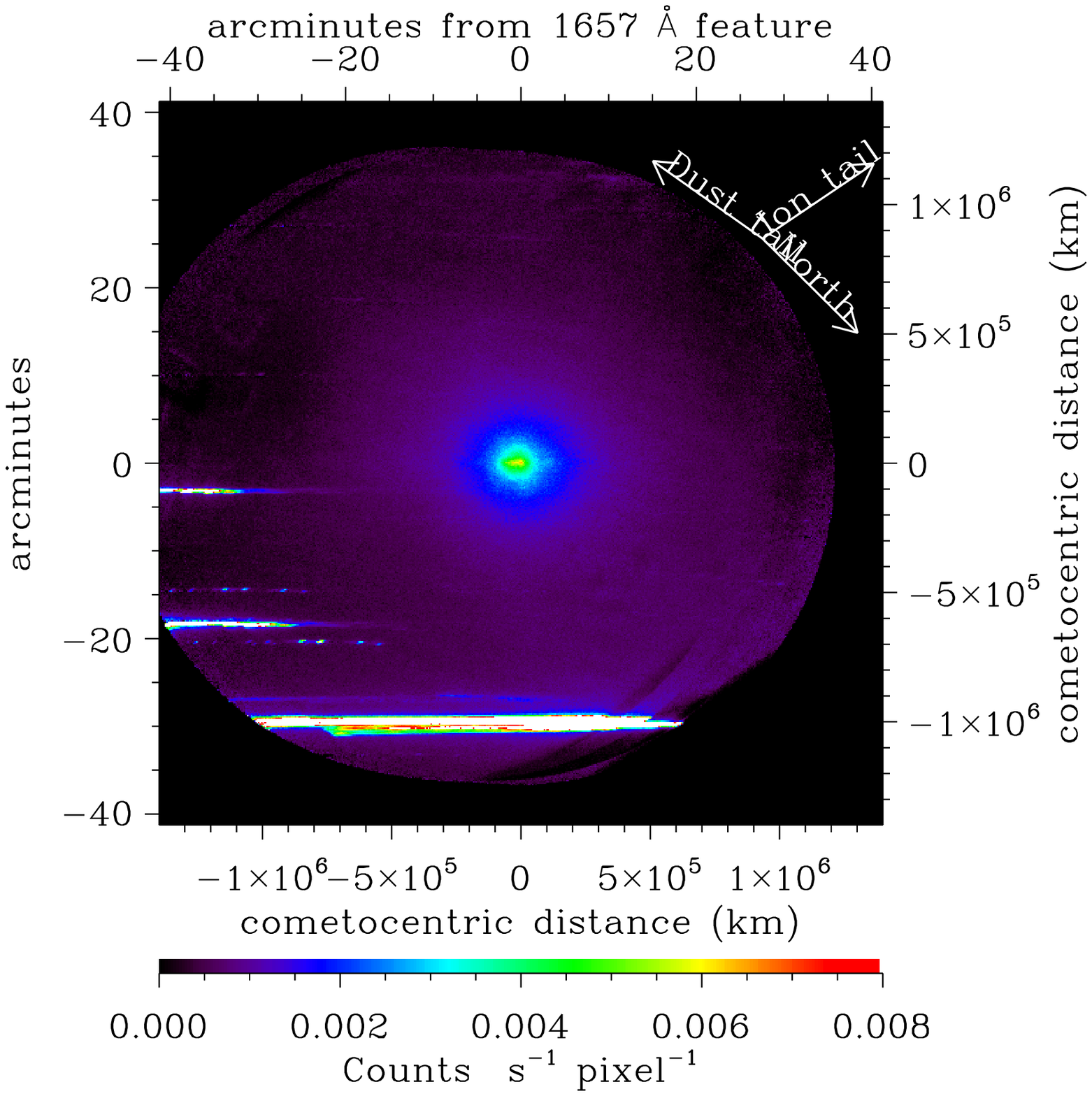}
  \caption[Machholz FUV Images]{Top left: \GALEX\ default pipeline
  processed direct-mode comet Machholz FUV image.  Top right: the
  same, but grism-mode orbit~1.  Bottom left (direct-mode) and right
  (grism-mode): images reconstructed in the frame of the comet, with
  background sources removed, and diffuse background subtracted, as
  described in \S\ref{reduce}.  Four grism-mode observations are
  co-added (note trailing of FOV).  The grism-mode images have been
  oriented so that the dispersion direction is horizontal and longer
  wavelengths are found to the right.  Note that three bright stars
  off the edge of the FOV have been dispersed by the grism into the
  FOV.}
  \label{fig:pipe_move_compare}
\end{figure}

The top two panels of \fig{pipe_move_compare} show images created by
the \GALEX\ pipeline processing system.  The 1657\,\AA\ multiplet is
the dominant feature at all distances from the nucleus.  Stars that
appear as point sources in the direct-mode image are streaks in the
grism-mode image since each stellar spectrum formed by the objective
grism is individually imaged on the detector.  On this scale, the
comet motion is not readily apparent, but must be considered in order
to provide high-quality radial profiles of the direct-mode image and
co-add all of the grism images.

\begin{deluxetable}{llllllll}
\centering
\tablewidth{0pt}
\tablecaption{Observational Parameters}
\tablehead{
  \colhead{Parameter} &
  \colhead{Value} &
}
\startdata
Comet observed		& 2005-03-01 \\
Heliocentric distance	& 1.3\,AU	\\
Heliocentric velocity	& 11.0\,\kms	\\
Heliographic lat.	& 30.3\arcdeg \\
Heliographic lon.	& 141.0\arcdeg \\
Geocentric distance	& 0.75\,AU \\
FOV R.A.		& 4\,hr 13\arcmin 50,\arcsec\\
FOV Dec.		& 82\arcdeg 45\arcmin 10\arcsec\\
\enddata
\label{tab:obs}
\end{deluxetable}

\begin{deluxetable}{lrrlllll}
\tabletypesize{\scriptsize}
\tablewidth{0pt}
\tablecaption{Log of \GALEX\ comet C/2004 Q2 (Machholz) observations used in this work}
\tablehead{\multirow{2}{*}{Type} & \multirow{2}{*}{Name\tablenotemark{a}} & \colhead{Exp} & \multirow{2}{*}{Date} & \colhead{Start}\\
  			  &					& \colhead{time (s)} &		 & \colhead{Time}
}
\startdata
Back. direct	& 1-fd-x.fits & 271 	& 2005-01-30 & 03:04:42\\
Comet direct	& 2-fd-x.fits & 1635	& 2005-03-01 & 19:29:19\\
Comet grism	& 1-fg-x.fits & 1618	& 2005-03-01 & 11:16:19\\
Comet grism	& 2-fg-x.fits & 1619	& 2005-03-01 & 12:54:57\\
Comet grism	& 4-fg-x.fits & 1619	& 2005-03-01 & 16:12:13\\
Comet grism	& 5-fg-x.fits & 1619	& 2005-03-01 & 17:50:51\\
\enddata
\tablenotetext{a}{filename prefixes are GI1\_104001\_COMET\_2004Q2\_000}
\label{tab:log}
\end{deluxetable}

\section{Data Reduction}
\label{reduce}
The \GALEX\ MAMA imagers detect and time tag individual photons.  The
photon lists are telemetered to the ground along with the spacecraft
aspect information.  Using reference stars in the FOV, the astrometric
solution is refined to better than 0.6\arcsec\ \citep{morrissey07}.
In the reference frame of the stars, comets are trailed during the
\about 1500\,s\ \GALEX\ observing segments (``visits'').  When several
orbits are available, comets appear multiple times across co-added
images.  To eliminate comet trailing and multiple comet images, we
used astrometrically corrected photon lists and an accurate comet
ephemeris to cast photons into a comet frame of reference, which
trails the stars.  We obtain high-precision comet ephemerides from the
JPL HORIZONS system using ``@galex'' as the observatory, which ensures
that effects such as \GALEX\ orbit parallax are properly considered.
All coordinate transformations were done using the IDLASTRO
routines.\footnote{http://idlastro.gsfc.nasa.gov/}

Comet motion across the \GALEX\ FOV and a mask that blocks out
background sources allows us to completely remove flux from identified
background sources.  We used direct-mode observations of the same star
field recorded a month before the Machholz arrived to create a high
quality background source mask.  Sources were identified by Emmanuel
Bertin's SExtractor
package\footnote{http://sextractor.sourceforge.net/} that is run as
part of the normal \GALEX\ pipeline.  

For the direct-mode observation, masks were created using the extreme
corners of the sources tabulated in the SExtractor catalog columns
XMIN, XMAX, YMIN, and YMAX.  The mask was used to blank out the
background sources in separate count and exposure time images that
were accumulated as the photon lists were read.  The mask was applied
every time the predicted position of the comet moved 1~pixel relative
to the background sources.  The \GALEX\ flat-field image was applied
to the exposure image in a similar way.  When the end of the photon
list was reached, the count image was divided by the exposure image to
produce a flat-fielded count rate image with the background sources
effectively erased.  The FOV of this image was extended in the
direction of comet motion in the same way that the stars were trailed
when we did not mask them.

For the grism-mode observations we converted the individual source
masks into rectangles elongated in the dispersion direction and
proceeded with the reduction in the same way as with the direct-mode
for each visit individually.  Photometrically correct addition of the
four grism visits was achieved by adding count and exposure-time
images and then dividing the results.  Because the grism can disperse
light from stars that are out of the FOV, our method of working from
sources cataloged in the direct-mode image did not always produce
masks for stars.  These stars could easily be masked by hand or by
modeling the comet emission (described in \S\ref{analysis}),
subtracting it from the comet image cast in the reference frame of the
stars by a procedure similar to which the stars are masked and
inspecting the residuals for sharp features.  For the purposes of this
work, stars left in the image did not prove to be problematic.

In order to assure that our method of processing the \GALEX\ photon
list was consistent with that of the default \GALEX\ pipeline, we ran
our software without the comet ephemeris and without masking the
stars.  We found minimal count rate differences between our images and
pipeline images, except within 100 1.5\arcsec\ pixels of the edge of
the circular FOV.  We attribute this to details in the flatfielding
process.  In order to avoid detector burn in, the \GALEX\ spacecraft
executed a 1.5\arcmin\ spiral motion during pointed observations
\citep{morrissey05}.  The default \GALEX\ pipeline applied the
detector flat in the detector coordinate system every few seconds
while images were being reconstructed in an astrometric coordinate
system.  The effect of the spiral motion on the flat-fielding at the
edge of the FOV was therefore correctly considered.  The flat-field
provided to guest observers by the \GALEX\ project is an average of
the detector flat over each orbit.  Our method of applying this flat
therefore did not take into consideration the spiral motion.  The flat
changes slowly as a function of position except at the edges, so the
differences in flat-fielding methods does not effect the quality of
the data except at the edges.  As the \GALEX\ images are 3840\by 3840
pixels, we simply masked out the 100~pixel wide annulus when we were
applying the flat-field corrections in our procedure.

The lower-left panel of \fig{pipe_move_compare} shows the resulting
direct-mode image with the background discussed in the next section
subtracted.  The removal of discrete sources has been highly
successful.  The lower-right panel of \fig{pipe_move_compare} shows
the corresponding grism-mode image, which is a co-add of 4 visits.

\subsection{Background}
\label{background}

Since the comet emission filled the field-of-view, it was not possible
to estimate the contribution of the background from the comet image
itself.  We used the background image recorded a month before the
comet image for this purpose.  Discrete sources were removed using the
mask generated from the SExtractor catalog.  In order to fill in the
missing data where the sources were removed, we smoothed the image
using a 100\by100 pixel boxcar average.  The resulting image was
translated to the average R.A.\ and DEC of the comet during the
direct-mode observation.

\begin{figure}
  \plotone{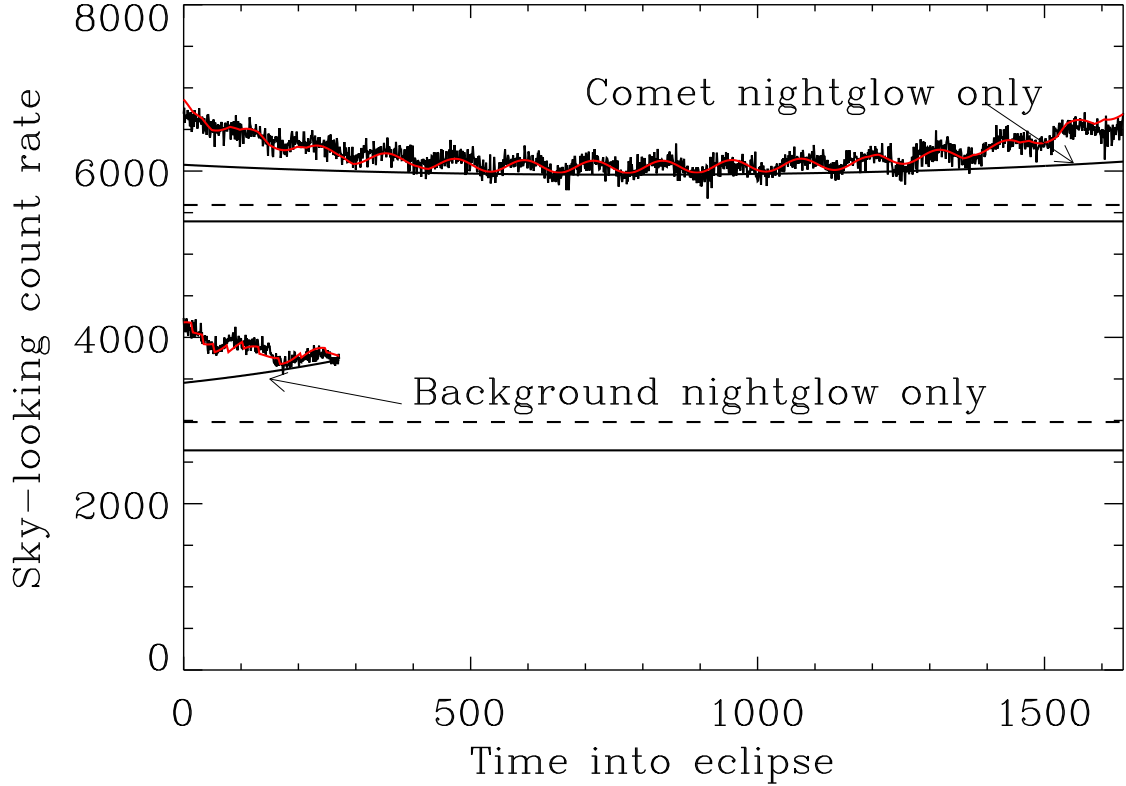}
  \caption[\GALEX\ FUV sky looking detector count rates]{Count rate in
  the \GALEX\ FUV detector for the comet (upper data points) and
  background (lower data points) visits.  The count rate has been
  corrected for detector effects such as hot spots.  The sinusoidal
  variation in the signals is due to a bright star near the edge of
  the FOV and the spiral motion of the \GALEX\ spacecraft.  The model
  described in the text plus the sinusoid is overplotted in red in
  both cases.  The model with $B$ = 1 (nightglow only) is overplotted
  to demonstrate the relative contributions of the day and nightglow
  signals.  The average astrophysical signal without airglow is
  plotted as the straight lines for the two cases described in the
  text: $\rho_0$ = 0.7 (solid) and $\rho_0$ = 0.5 (dashed).}
  \label{fig:sky_looking}
\end{figure}

Before subtracting the image from the data, we checked for any effects
that might contribute to variation of the background level over the
course of one month.  \citet{sujatha09}, in studying multiple \GALEX\
observations of the Sandage nebulosity show that in the FUV, airglow
from the 1356\,\AA\ and possibly 1304\,\AA\ lines is responsible for
the primary variation in the total \GALEX\ count rate as a function of
time over short and long timescales.  \fig{sky_looking} shows the
count rate in the \GALEX\ FUV detector for the comet (upper curve) and
background (lower curve) after removal of detector effects such as hot
spots.  The sinusoidal variation is induced by the combination of a
bright star near the edge of the FOV and the spiral motion of the
\GALEX\ spacecraft.  Otherwise, the count rate curves resemble the
parabol\ae\ described by \citet{sujatha09}.  We adapted the their
method to estimate the \textit{difference} between the airglow
contribution in the comet image and the background image:
-5.66$\times$10$^{-5}$\,counts\,s$^{-1}$\,pixel$^{-1}$.  The primary
contribution to the difference in the airglow contribution came from
the fact that the background image was recorded only on the steep part
of the count rate curve, so its average airglow was high compared to
the comet observation, which covers the entire parabola.

We note that the \citet{sujatha09} method is largely empirical and
must be repeated for each observing geometry.  The comet and the
Sandage nebulosity were both high-dec sources, so we expect our use of
their parabol\ae\ to be reasonably accurate.  However, the sensitively
of our lifetime measurement to the precise value used for the
background motivated us to investigate the airglow effect more
carefully.  We demonstrate here that a simple atmospheric model,
geometric considerations of earth shadow on the atmosphere and
accurate consideration of observing geometry, explains the variation in
\GALEX\ count rate seen during each eclipse period.  Much like
photometric data taken at various zenith distances, our method can be
used to independently remove the effects of airglow from \textit{any}
\GALEX\ eclipse.

We constructed a model atmosphere using an exponential rooted in the
F-region (altitude 300\,km) with an initial density $\rho_0$ (detected
\GALEX\ FUV counts per km at an altitude of 300\,km) and a scale
height, $h$.  We assumed that the distribution was the same over every
point on the Earth but we defined a unitless ``day boost'' parameter,
$B$, which multiplied the density of emitters when they were in
sunlight.  We used the orbit of \GALEX\ described by two line elements
(TLEs) archived at http://www.spacetrack.org to determine the position
of the satellite as a function of time, a solar ephemeris provided by
the SUN-POS.PRO module of the IDLASTRO routines, and a simple stepwise
integrator to compute the airglow brightness for each line of sight as
a function of time into the visit.  The final component of the model
was the astrophysical count rate, which we assumed to be constant over
the visit.  The red solid lines in \fig{sky_looking} show the
resulting best-fit model, including the sinusoid discussed above.  Our
model gave good fits for a range of parameters corresponding to
airglow \textit{differences} of
--5.57$\times$10$^{-5}$\,counts\,s$^{-1}$\,pix$^{-1}$ to
--3.6$\times$10$^{-5}$\,counts\,s$^{-1}$\,pix$^{-1}$.  An interesting
extension to this exercise would be to use more sophisticated
ionospheric and plasmaspheric models to derive the airglow emitter
profiles and increase the accuracy of the \textit{absolute} airglow
determination.

Having derived a best-fit background image, we subtracted it from the
comet image.  There was a slight oversubtraction around the edges due
to the background image smoothing and the elongation of the comet
image.  The effected points were removed with an annular mask
175~pixels wide.  Residual emission around the hot spot in the
southwest portion of the image and a bright source toward the west
were also masked.  The resulting image is shown in the lower-left
panel of \fig{pipe_move_compare}.  The co-added grism-mode images
formed by our reduction procedure are shown in the lower-right panel
of \fig{pipe_move_compare} with diffuse background determined in
\S\ref{analysis} subtracted.

\section{Analysis}
\label{analysis}

We used the combination of \GALEX\ direct and grism-mode data to
measure the carbon ionization lifetime.  We began by assuming that
\ion{C}{1} emission dominates the passband at all cometocentric
distances and that CO is the sole parent molecule.  As discussed by
\citet{huebner92} and \citet{tozzi98}, CO dissociates in significant
amounts into both ground state carbon, C($^3$P) and the metastable
state \CID.  \CID\ decays to C($^3$P) in 4077\,s (a rate of
2.45$\times$10$^{-4}$\,s$^{-1}$), independent of heliocentric distance
\citep{hibbert93}.  \CID\ can be photoionized with a predicted rate
that is \about10 times faster than that of C($^3$P) \citep{huebner92}.
However, the resulting photoionization lifetime, 2.8$\times$10$^5$\,s
(rate of 3.6$\times$10$^{-6}$\,s$^{-1}$), is still long (rate is
slow) compared to the metastable lifetime (rate), so for the purposes
of this work, we neglect the time spent in the metastable state and
assume all of the CO dissociates directly to C($^3$P).  In
\S\ref{haser}, we explore the differences in the photochemical
ejection velocities of C($^3$P) and \CID, which could, in principle,
affect our results, and find them to be small (\about25\%).

\begin{figure}
  \plotone{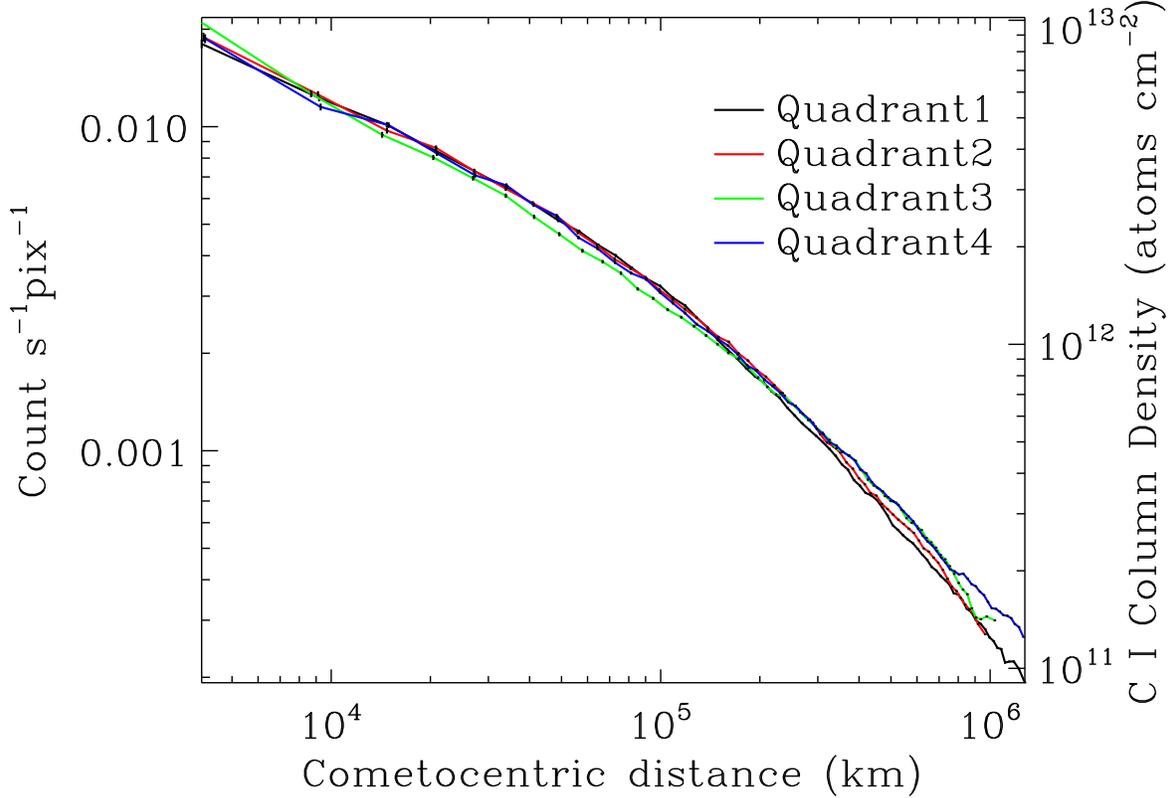}
  \caption[Machholz \ion{C}{1} profiles]{Quadrant-by-quadrant radial
  profiles of comet Machholz FUV emission.  The left-hand Y-axis shows
  the background-subtracted average count rate per pixel in each
  radial bin.  The right-hand Y-axis shows the count rate converted
  into \ion{C}{1} column density using the $g$-factors described in
  the text and the assumption that all of the emission in the FUV band
  is from \ion{C}{1}.  Quadrant~4 is centered on the antisunward or
  ``Dust tail'' direction, as indicated in the lower-left panel of
  \figp{pipe_move_compare}.  Quadrant~1 is counterclockwise from
  quadrant~4, etc.  Statistical error bars (total vertical extent
  2--$\sigma$) are shown and are comparable to the line widths.}
  \label{fig:profiles}
\end{figure}

We fit two-component \citeauthor{haser57} models to radial profiles
formed from our background-subtracted direct-mode image (\ref{haser},
\figp{profiles}).  We did this as a function of airglow offset in
order to estimate the effect of our background subtraction uncertainty
on the carbon lifetime (\figp{TCI_back}).  We find that the lifetime
measured by this method agrees well with the theoretical carbon
ionization lifetime calculated using values for solar wind and
photoionization tabulated by \citet{rubin09} assuming comet Machholz
was in the slow solar wind (\figp{TCI_back}, \tab{rates}).

%
%
%

\fig{profiles} shows radial profiles of four quadrants of the
background-subtracted direct-mode comet image shown in the lower-left
panel of \figp{pipe_move_compare}.  Quadrant~4 is centered on the dust
tail, quadrant 1 is oriented 90\arcdeg\ counter-clockwise from
quadrant~4, etc.  The profiles were created using an adaptive binning
ring-sum algorithm similar to that described by \citet{harris02}.  For
the profiles shown in \fig{profiles}, at least 1000 counts per radial
bin were used.  For our fitting exercises, there were at least 10,000
counts per bin.  The profiles show good agreement except for the one
oriented toward the dust tail.

\subsection{The \citeauthor{haser57} model}
\label{haser}

\citet{haser57} first solved the simple differential equation which
describes the radial distribution of a species traveling radially away
from a comet nucleus at velocity $v_{bulk}$ and with lifetime against
destruction, $\tau_p$.  A similar differential equation can be solved
for the case of a daughter species, resulting in the 2-component
\citeauthor{haser57} model \citep[e.g.,][]{krishna-swamy86}.  A
critical component of the 2-component \citeauthor{haser57} model is
the ejection velocity, $v_e$, of the daughter from the parent
molecule.  \citet{festou81a} pointed out that the common usage of
$v_e$ in the \citeauthor{haser57} model was unphysical: the daughters
were frequently assumed to be ejected radially outward from their
parents, when in fact, they would be emitted isotropically from the
point of destruction of the parent.  The vectorial model of
\citet{festou81a} addresses this shortcoming.  Many workers have noted
that the 2-component \citeauthor{haser57} model provides good fits to
comet profiles of daughter species.  The problem is mapping the input
parameters of the \citeauthor{haser57} model (e.g., parent and
daughter outflow velocities) to independently determined quantities
(e.g., parent outflow velocity and daughter ejection velocity).
\citet{combi04} provide formul\ae\ for this mapping that allow the
computationally simple \citeauthor{haser57} model to be used to
emulate the more sophisticated models.  We use the \citet{combi04}
formul\ae\ in all our calculations.

As outlined above, our starting point was to assume CO is the sole
parent of \ion{C}{1}.  We also assume CO is produced within the
collision sphere and thermalized to the bulk outflow velocity,
$v_\mathrm{bulk}$, of the coma.  Preliminary analysis of our
contemporaneously recorded \GALEX\ NUV data suggests that comet
Machholz had a water production rate in the log(Q(\water)) = 28 -- 29
range.  According to the formul\ae\ given by \citet{budzien94} and
\citet{tseng07}, this translates into $v_\mathrm{bulk}$ = 0.74\,\kms\
and 1.02\,\kms, respectively at the comet's heliocentric position.
Combining the \citet{huebner92} CO photodissociation rate and
\citet{rubin09} CO solar wind ionization rates results in a total
solar quiet CO lifetime of 7.9$\times$10$^5$\,s (rate of
1.3$\times$10$^{-6}$\,s$^{-1}$).  \citet{tozzi98} use an improved set
of calculations and find a solar maximum CO photodissociation lifetime
of 4$\times$10$^5$\,s (2.5$\times$10$^{-6}$\,s$^{-1}$).  We estimate
the solar minimum CO lifetime to be $\lesssim$~5$\times$10$^5$\,s
($\gtrsim$~2$\times$10$^{-6}$\,s$^{-1}$).

\begin{figure}
  \plotone{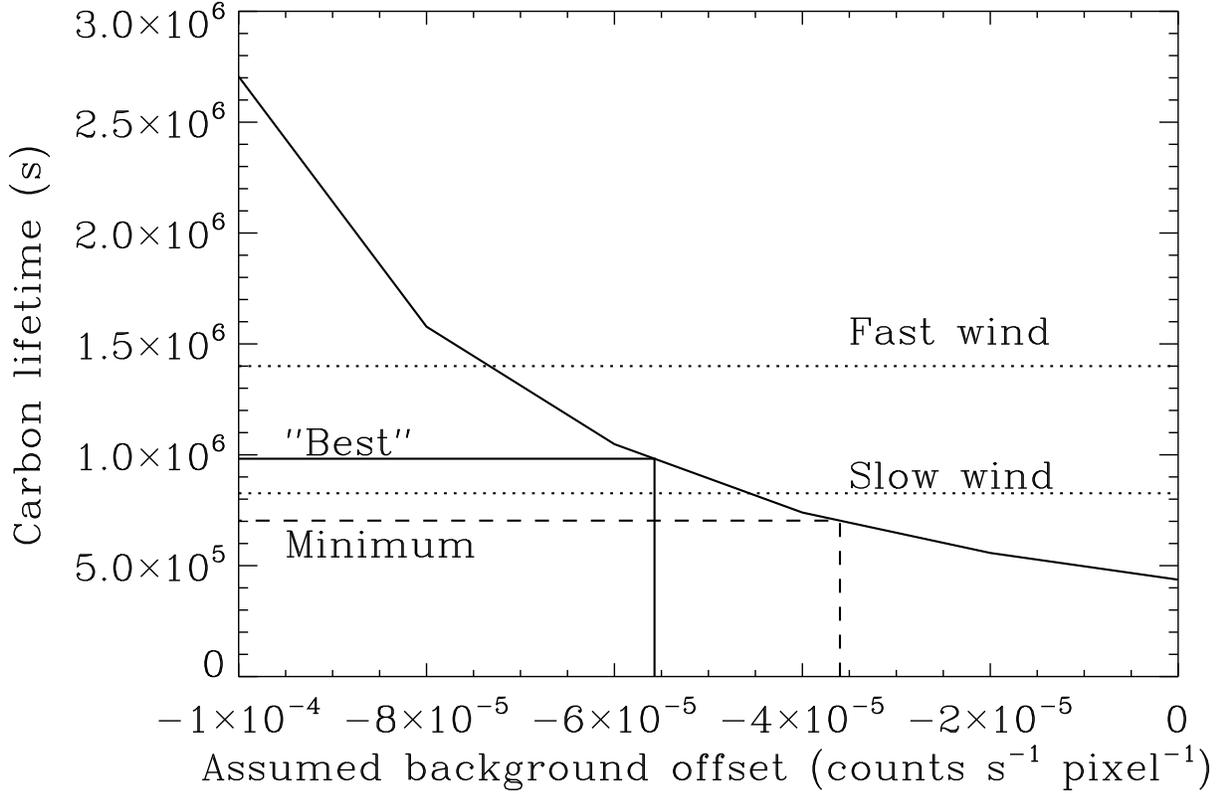}
  \caption[Carbon lifetime vs.\ assumed background]{Carbon lifetime at
  1\,AU vs.\ assumed background offset between the comet and
  background direct-mode images.  The best-estimate value of our
  background offset calculated in \S\ref{background} to be
  --5.57$\times$10$^{-5}$\,counts\,s$^{-1}$\,pixel$^{-1}$ is shown as
  the solid vertical line.  The resulting carbon lifetime is similarly
  indicated graphically and labeled ``Best.''  The minimum background
  offset is indicated in this way with the dashed lines.  The
  calculated values of the carbon lifetime from the first two columns
  of \tab{rates} are shown as dotted lines.  The data indicate that
  during our observations, comet Machholz was embedded in the slow
  solar wind.}
  \label{fig:TCI_back}
\end{figure}

As discussed above, we can neglect ionization out of the \CID\ state
because the metastable lifetime is so much shorter than the ionization
lifetime, however, we must check to see if the ejection velocities,
$v_\mathrm{e}$, of C($3$P) and \CID\ from CO are sufficiently
different to require individual consideration.  \citeauthor{huebner92}
calculate $v_\mathrm{e}$ = 4.9\,\kms and 3.9\,\kms, respectively.
\citeauthor{tozzi98} find the ejection velocity, $v_\mathrm{e}$, of
C($3$P) from CO to be 4\,\kms.  \citeauthor{tozzi98} do not directly
calculate $v_\mathrm{e}$ values for \CID, but rather adopt a range of
values from $v_\mathrm{bulk}$ to 3.9\,\kms.  We used a less
conservative lower limit and assumed that the \CID\ $v_\mathrm{e}$
value is within 25\% of the C($3$P) $v_\mathrm{e}$ value, as suggested
by the \citeauthor{huebner92} calculations.  As discussed below and
shown in \fig{TCI_back}, this is comparable to the uncertainty in our
fitting and background removal activities, so we neglect it.

The formula of \citet{whipple76} shows that the collision sphere, over
which gas interactions transfer excess energy between gas species, was
\about 2500\,km for Machholz.  This is much smaller than scale lengths
of CO or any of the major carbon parent molecules.  Thus, the motion
of carbon is dominated by its ejection velocity.  Since $v_\mathrm{e}$
is significantly larger than $v_\mathrm{bulk}$, the \citet{combi04}
formul\ae\ show that the daughter lifetime that would be derived from
sophisticated models is essentially unaltered from that which is
derived from the \citeauthor{haser57} model.  Conversion of our
measured \citeauthor{haser57} scale lengths to the \ion{C}{1} lifetime
is a simple matter of dividing by the assumed ejection velocity.

We used MPFIT \citep{more78, markwardt09} to fit two-component
\citeauthor{haser57} models to the profiles.  Because the statistical
errors on the profiles we were fitting were so small, the formal
errors of all the parameters were always less than 1\%.  The
off-diagonal elements of the correlation matrix were also always less
than 1\%.  The two-component \citeauthor{haser57} model was able to
accurately reproduce all of the profiles in \fig{profiles} except for
the inner-most point, which consistently fell above the model.

We experimented with excluding data below small cometocentric
distances, $r_\mathrm{min}$ to see if this effected the \ion{C}{1}
lifetime.  As $r_\mathrm{min}$ was increased from 0 to
1$\times$10$^5$\,km, there was a 20\% decrease in the measured
\ion{C}{1} lifetime.  Finding this to be an acceptable range of
uncertainty in the fitting exercise, we choose the midpoint,
5$\times$10$^4$\,km as our best value and proceeded with the other
experiments described below.

Next, we varied $r_\mathrm{max}$, the upper cutoff of the data we fit,
and the assumed airglow offset value.  We created contour plots of the
\ion{C}{1} lifetime in the plane of airglow offset versus
$r_\mathrm{max}$ for a variety of combinations of quadrants.  We found
that quadrant~1 resulted in the most consistent behavior as a function
of $r_\mathrm{max}$.  For all $r_\mathrm{max} > $ 3$\times$10$^5$\,km,
in quadrant~1 all the measured \ion{C}{1} lifetime values fell within
10\% of each other.  This is important.  It means that the \GALEX\
FOV, which extends to \about1.1$\times$10$^6$\,km is large enough to
reliably measure the carbon scale length, even though at this
heliocentric distance, it was larger than the \GALEX\ FOV.

\fig{TCI_back} summarizes our fitting exercises.  The \ion{C}{1}
lifetime is plotted against the assumed offset between the comet and
background airglow levels.  The plot was made by fitting quadrant~1
data between $r_\mathrm{min}$ = 5$\times$10$^4$\,km and
$r_\mathrm{max}$ = 1.1$\times$10$^6$\,km (the largest radial bin
containing at least 10,000 counts).  It is at the edge of the radial
distribution that the background subtraction accuracy becomes
particularly important, as any over/undersubtraction is directly
reflected as a change in the number density of CI and therefore the
scale length.  We mark the two values of the airglow offset level
discussed in \S\ref{background} with the solid and dashed lines.  The
resulting 1\,AU \ion{C}{1} lifetimes are 7.1 and 9.6$\times$10$^5$\,s
for the ``best'' and minimum cases, respectively.  These correspond to
total ionization rates of 1.4 and 1.0$\times$10$^{-6}$\,s$^{-1}$,
respectively.  \tab{rates} summarizes these results.

\subsection{Comparison to physical processes}
\label{compare}
Photoionization is generally assumed to be the dominant ionization
process for most cometary molecules and radicals, however, for
long-lived species such as carbon and oxygen, solar wind ionization
processes become non-negligible \citep[e.g.,][]{axford72, rubin09}.
We find that for carbon, solar wind ionization processes are
comparable to photoionization.  \tab{rates} summarizes calculations of
carbon ionization lifetimes (rates) from the dominant physical
processes for a variety of solar conditions.  We observed comet
Machholz near the end of solar cycle\,23, so we assume quiet sun
conditions from the perspective of the photon field.  The heliographic
latitude of Machholz was 30\arcdeg, so Machholz was likely on the
upper edge of the slow solar wind zone identified by \Ulysses\ and
\SOHO\ SWAN observations \citep{smith_marsden95, bzowski03}.  Our
measured ionization lifetime of 7.1 -- 9.6$\times$10$^5$\,s (1.0 --
1.4$\times$10$^{-6}$\,s$^{-1}$) therefore compares favorably with the
predicted lifetime of 8.2$\times$10$^5$\,s
(1.2$\times$10$^{-6}$\,s$^{-1}$).  In \fig{TCI_back}, we present a
graphical interpretation of these results combined with the
presentation of our most easily identified systematic uncertainty:
determination of the relative contributions of airglow in the comet
and background images.

\begin{deluxetable}{lcccl}
\tablewidth{0pt}
\tablecaption{Carbon Ionization Lifetimes (Rates) at 1\,AU}
\tablehead{
  \multirow{3}{*}{Process} & \multicolumn{3}{c}{{Ionization Lifetime $\times$10$^5$\,s (Rate $\times$10$^{-6}$\,s$^{-1}$)}} & \multirow{3}{*}{Reference} \\
  & \colhead{Quiet Sun} & \colhead{Quiet Sun} & \colhead{Active Sun} \\
  & \colhead{Slow Wind\tablenotemark{a}} & \colhead{Fast Wind\tablenotemark{b}} & \colhead{Slow Wind\tablenotemark{a}}
}
\startdata
C $+\,h\nu \rightarrow$ C$^+ + e^-$	& 24 (0.41) & 24 (0.41)	& 10 (0.92)	& \citet{huebner92}\\
C $+$ H$^+ \rightarrow$ C$^+ + $ H	& 17 (0.59) & 40 (0.25)	& 17 (0.59)	& \citet{rubin09}\\
C $+\, e^- \rightarrow$ C$^+ + 2e^-$	& 48 (0.21) & 20 (0.05)	& 48 (0.21)	& \citet{rubin09}\\
\hline
Total predicted				& 8.2 (1.21) & 14 (0.71)& 5.8 (1.72)	\\
Measured 				& \multicolumn{2}{l}{7.1 -- 9.6 (1.0 -- 1.4)} \\
\enddata
\tablenotetext{a}{$v$ = 400\,\kms, $n_e$ = 10\,cm$^{-3}$}
\tablenotetext{b}{$v$ = 750\,\kms, $n_e$ = 2.5\,cm$^{-3}$}
\label{tab:rates}
\end{deluxetable}


\subsection{Grism Image analysis}
\label{grism}

In \S\ref{haser} we described \citeauthor{haser57} model fits to the
direct-mode profiles and showed that the measured \ion{C}{1} lifetime
was relatively stable as a function of $r_\mathrm{min}$ and
$r_\mathrm{max}$ (20\%).  However, these fits all required a parent
scale length that was extremely short, indicating the presence of
contamination from the shorter-lived species in the \GALEX\ FUV
bandpass.  We use a preliminary analysis of the \GALEX\ grism-mode
data to identify the species contributing the contaminating emission
and demonstrate that this contamination is unlikely to effect our
\ion{C}{1} lifetime measurement.

The general method we use for analyzing the grism-mode images is
forward modeling.  We create a radial profile of the expected emission
in a particular line or spectroscopically unresolved band.  Our
software is structured so that any model profile can be used, but for
these analysis, we use the 1 or 2-component \citeauthor{haser57}
model, as appropriate, with scale lengths modified by the formul\ae\
of \citet{combi04} to emulate the output of more sophisticated coma
models.  The profile is scaled by the product of the production rate,
the $g$-factor of the feature in question (see below and
\tab{g_factors}) and the effective area of the \GALEX\ instrument at
the wavelength of the feature
\citep{morrissey07}\footnote{http://galexgi.gsfc.nasa.gov/docs/galex/Documents/PostLaunchResponseCurveData.html}.
We then create a figure of rotation from the profile, thus
transforming the profile into a coma image.  We offset the image
according to the \GALEX\ grism dispersion relation from the
astrometric center of the comet, determined by our reduction procedure
described in \S\ref{reduce}.  Next, we convolved the image by an
estimate of the \GALEX\ point spread function in grism-mode, which we
create by making a figure of rotation from the cross-dispersion
profile model generated from stellar spectra in the grism-mode image
by the \GALEX\ pipeline processing system.  We then repeated the
process of converting the model profile into a coma image for each of
the three primary \GALEX\ grism orders and summed the results.

We calculated $g$-factors at 1\,AU and a heliocentric velocity of
11\,\kms\ using a high-resolution solar spectrum from the Solar
Maximum Mission normalized to absolution fluxes measured by \SOLSTICE\
together with absorption oscillator strengths available from NIST.
For the purposes of these calculations, we assumed optically thin
conditions.  The implications of this assumption, particularly for CO,
are discussed below.  The CO $g$-factors were calculated assuming an
excitation temperature of 70\,K.  The $g$-factors and \GALEX\ grism
mode effective areas are listed in \tab{g_factors}.

\begin{deluxetable}{lccccl}
\tablewidth{0pt}
\tablecaption{$g$-factors at 1\,AU}
\tablehead{
  \colhead{Feature} & \colhead{$g$-factor\tablenotemark{a}} & \colhead{Eff. Area\tablenotemark{b}}}
\startdata
\ion{C}{1} 1561\,\AA & 0.93	 & 18.9\\
\ion{C}{1} 1657\,\AA & 4.22	 & 11.5\\
CO total	     & 0.16	 & 14.9 \\
\ion{S}{1} 1425\,\AA & 0.29	 & 8.9\\
\ion{S}{1} 1475\,\AA & 0.25	 & 21.6\\
\ion{S}{1} 1813\,\AA & 4.90 	 & 2.5\\
\enddata
\tablenotetext{a}{$\times$10$^{-5}$ phot s$^{-1}$ particle$^{-1}$}
\tablenotetext{b}{Grism-mode peak order, cm$^2$}
\label{tab:g_factors}
\end{deluxetable}

%

\begin{figure}
  \plottwo{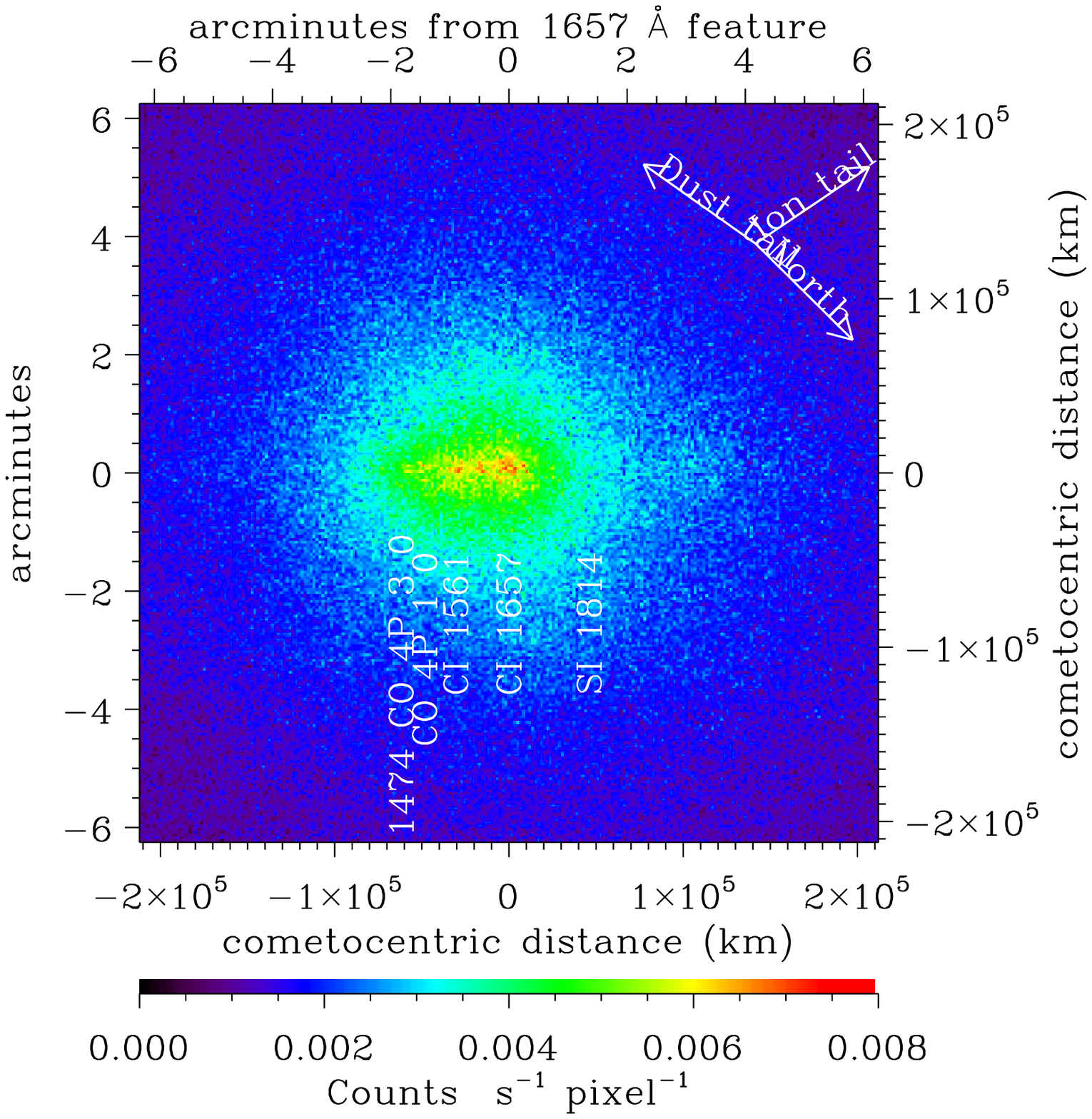}{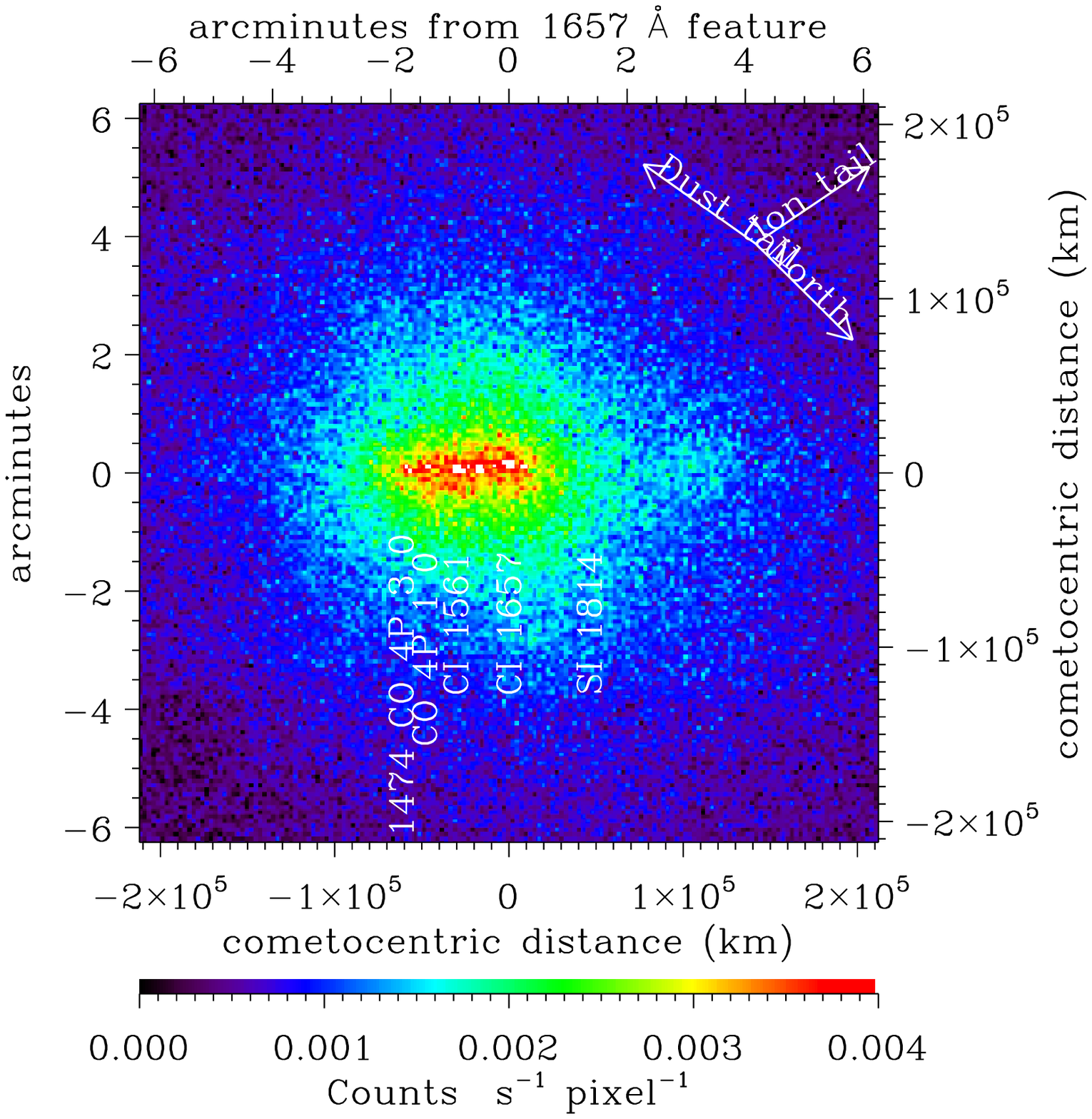}\\
  \plottwo{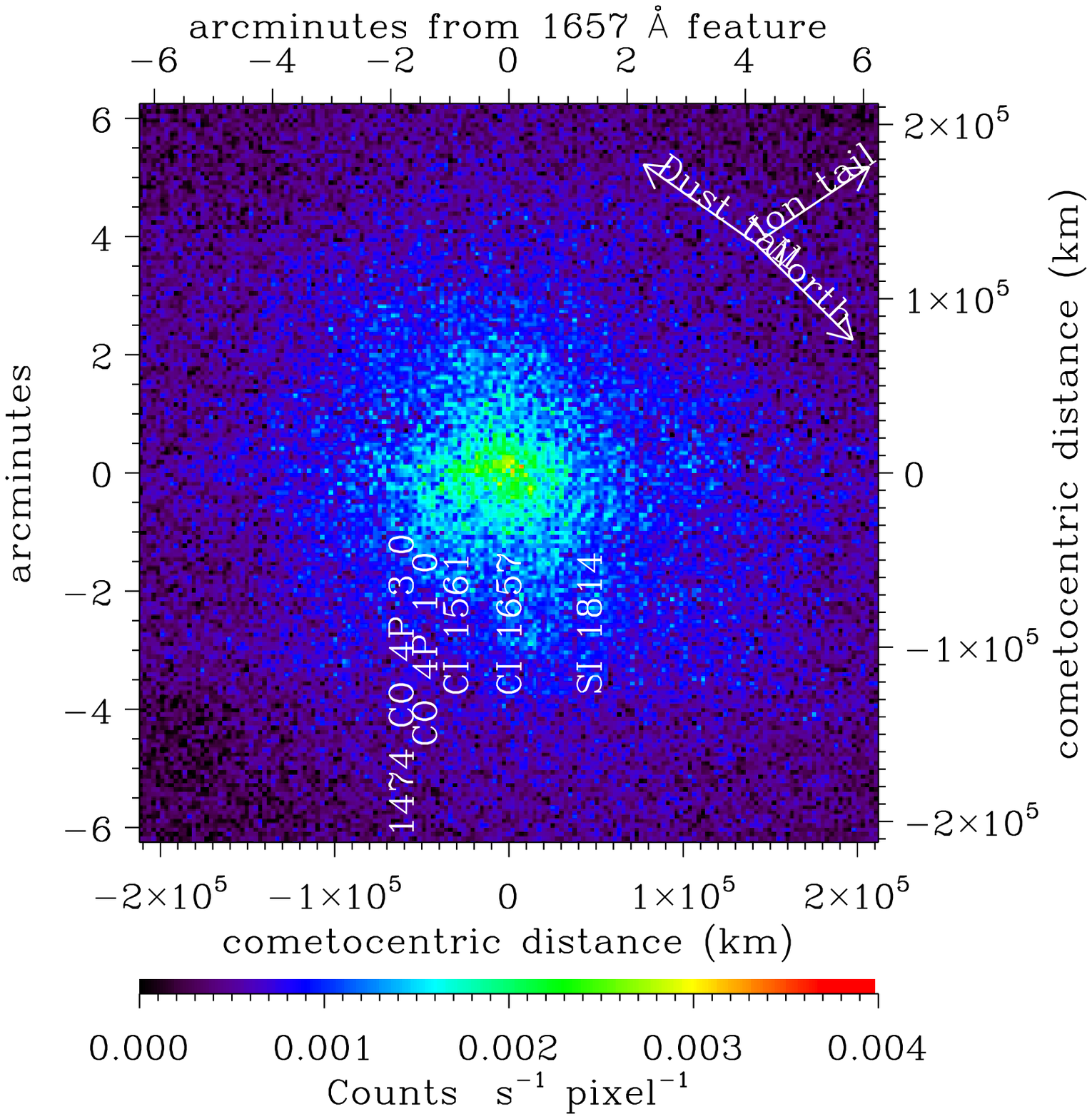}{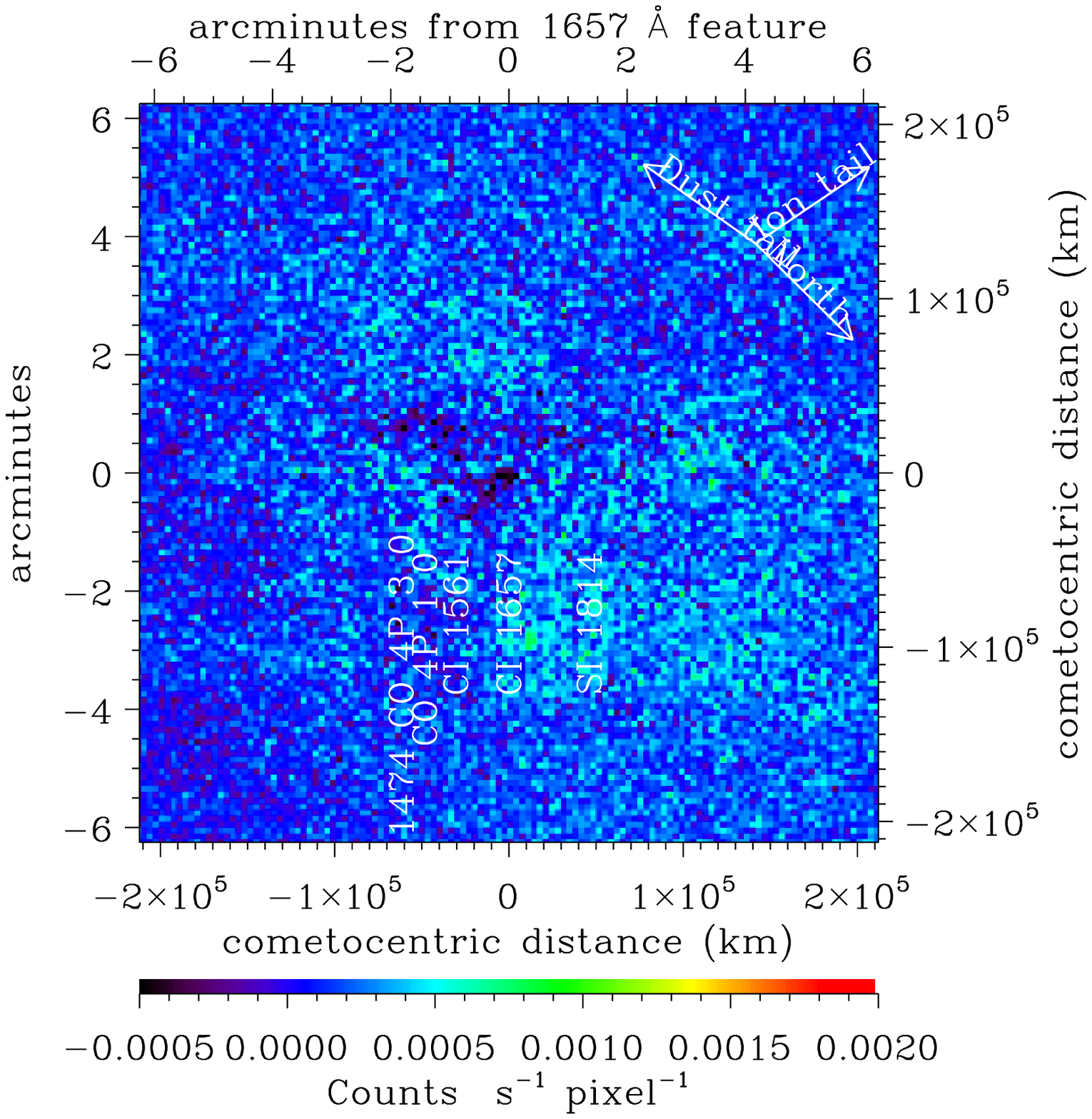}
  \caption[Machholz FUV grism-mode analysis]{Machholz FUV
  grism-mode analysis.  The frames show the inner
  2$\times$10$^{5}$\,km of the co-added comet Machholz \GALEX\
  grism-mode images in various stages of residual subtraction.
  Background described in the text has been subtracted from all
  frames.  The primary grism order positions of the brightest features
  in the FUV bandpass have been marked.  Top left: no model
  subtraction.  Top right: subtract \ion{C}{1} from a CO parent with
  Q(C) = 2.8$\times$10$^{28}$\,s$^{-1}$.  Bottom left: also subtract
  CO, with the profile described in the text and Q(CO) =
  3.0$\times$10$^{28}$\,s$^{-1}$.  Bottom right: modify original Q(C)
  to 1.8$\times$10$^{28}$\,s$^{-1}$, Q(CO) =
  2.8$\times$10$^{28}$\,s$^{-1}$ and subtract Q(C) =
  1.2$\times$10$^{28}$\,s$^{-1}$ from CH$_4$.  To increase the signal
  to noise ratio in the residual images, they have been binned by
  3$\times$3, 3$\times$3, and 4$\times$4 pixels, respectively.  Also
  note changes in color bar limits.}
  \label{fig:grism}
\end{figure}

We determined the background level in our grism-mode data by
subtracting a preliminary model of the \ion{C}{1} emission based on
the direct-mode analysis.  We assumed that CO was the sole parent of
\ion{C}{1} and used the same \ion{C}{1} production rate, Q(C) =
2.8$\times$10$^{28}$\,s$^{-1}$, found in our direct-mode analysis.
After experimenting with Q(C), and the \ion{C}{1} scale length, we
found that the grism-mode background was much higher than expected
assuming the value from the direct-mode image and the ratio between
the total efficiencies of the two modes.  We attribute the high
background to emission from the unmasked sources discussed in
\S\ref{reduce}.  We found a best-fit background level of
5.4$\times$10$^{-4}$\,counts\,s$^{-1}$\,pixel$^{-1}$.  The top two
panels of \fig{grism} shows the inner 2$\times$10$^{5}$\,km of our
\GALEX\ grism-mode data before (left) and after (right) the \ion{C}{1}
subtraction.  The positions of the major features in the primary
\GALEX\ FUV grism order have been indicated.

After subtraction of the \ion{C}{1} emission, a pseudo-continuum was
evident (upper right-panel of \figp{grism}).  Since dust emission near
the nucleus is negligible in the FUV, we took this to be emission from
the more than 80 lines of the CO Fourth Positive system that fall in
the \GALEX\ FUV bandpass.  Most of these lines have been resolved and
well studied by previous high resolution long-slit spectroscopic
observations \citep[e.g.,][]{mcphate99, lupu07}.  \citeauthor{lupu07}
model the CO Fourth Positive emission as a function of cometocentric
distance to 3,000\,km with and without considering self-absorption
(optical depth) effects.  They find evidence for detectable optical
depth effects in the inner \about 1000\,km of comets with production
rates similar to that of Machholz.  Perhaps not surprisingly, when we
subtracted a one-component \citeauthor{haser57} model of the CO
emission from our \ion{C}{1} residual, there was obvious
oversubtraction at cometocentric distances $<$10,000\,km.  We emulated
optical depth effects by using a 2-component \citeauthor{haser57}
model with an effective parent lifetime of 10,000\,s (rate of
1$\times$10$^{-4}$\,s$^{-1}$) at 1\,AU.  The resulting residual,
assuming Q(CO) = 3.0$\times$10$^{28}$\,s$^{-1}$ is shown in the
lower-left panel of \fig{grism}.  Since we assumed optically thin
conditions, our fit is not sensitive to whether or not CO comes from a
nucleus-native source or is formed from a distributed source
\citep[e.g.,][]{eberhardt87, disanti01, brooke03, lupu07,
bockelee-morvan10}.  We look forward to further work on this subject
as we incorporate radiative transfer calculations in our CO profile
calculations and improve our grism background source cleaning
technique.

After subtracting the primary \ion{C}{1} and CO sources, residual
emission, centered roughly on \ion{C}{1} 1657\,\AA\ was evident.  This
was not an artifact of the unexpectedly large grism background
discussed above.  It is reasonably well described by \ion{C}{1} from a
parent with an effective 1\,AU lifetime of 1$\times$10$^5$\,s (rate of
1$\times$10$^{-5}$\,s$^{-1}$), assuming CO has a lifetime of
5$\times$10$^5$\,s (2$\times$10$^{-6}$\,s$^{-1}$; see \S\ref{haser}).
We experimented with various CO lifetimes ranging from
2$\times$10$^5$\,s (which happened to describe a large section of our
data well) to 7.9$\times$10$^5$\,s \citep{rubin09}.  These CO
photodissociation lifetimes correspond to rates of 5 and
1.3$\times$10$^{-6}$\,s$^{-1}$, respectively.  The problem with a very
short CO lifetime is that if all of the \ion{C}{1} came from CO, the
model clearly over-subtracted inside of 2$\times$10$^4$\,km.  From
this we deduced that the dominant carbon parent molecule cannot have a
lifetime as short as 2$\times$10$^5$\,s (5$\times$10$^{-6}$\,s$^{-1}$)
and that at least one more carbon parent molecule was needed.

Knowing that we needed at least one other carbon parent molecule
besides CO to provide \ion{C}{1} emission inside 2$\times$10$^5$\,km,
we turned to infrared observations of comet Machholz to provide
suggestions.  Based on the work of \citet{bonev09} and
\citet{kobayashi09}, CH$_4$ and CH$_3$OH were the most likely
candidates.  Our estimates of the maximum production rates of atomic
carbon from CO and these molecules is shown in \tab{parent}.  Of
particular note is our adoption of the \citet{huebner92} CO branching
ratios even though we use the \citet{tozzi98} CO lifetime.  According
to \citeauthor{huebner92}, a significant amount of CO is lost to
ionization, leaving only 50\% for \ion{C}{1} emission.
\citeauthor{tozzi98} find a significantly shorter total CO lifetime
than \citeauthor{huebner92} but do not calculate rates into the
various branches.

\begin{deluxetable}{llclcc}
\tablewidth{0pt}
\tablecaption{Carbon Parent Molecule Candidates}
\tablehead{
  \colhead{Parent} & \multirow{2}{*}{\% CO}  & \colhead{Q(C)$_\mathrm{max}$}& \colhead{Q(C)}& \colhead{$\tau_p$}& \colhead{$v_\mathrm{e}$}\\
  \colhead{Molecule}  & & \multicolumn{2}{c}{($\times$10$^{28}$\,s$^{-1}$)}	& \colhead{($\times$10$^5$\,s)} & \colhead{(\kms)}}
\startdata
CO\tablenotemark{a}& 100		& \nodata		& 2.8\tablenotemark{a}	& \nodata		& \nodata \\
CO		& 50\tablenotemark{b}	& 1.4			& 1.8			& $\lesssim$5\tablenotemark{c}	& 4\tablenotemark{c} \\
CH$_4$		& 35\tablenotemark{d}	& 1.0			& 1.2\tablenotemark{e}	& 1.4			& $>$3.1\\
CH$_3$OH	& 50\tablenotemark{d}	& 1.4			& \nodata		& $>$5.9		& 4.5\\
C\tablenotemark{f}& \nodata		& \nodata		& 2.8\tablenotemark{f}	& 8\tablenotemark{g}	& \nodata
\enddata
\tablenotetext{a}{CO Fourth Positive emission}
\tablenotetext{b}{\citet{huebner92}}
\tablenotetext{c}{See \S\ref{haser}}
\tablenotetext{d}{\citet{bonev09} and \citet{kobayashi09}}
\tablenotetext{e}{Decreases to 1.0$\times$10$^{28}$\,s$^{-1}$
  assuming CO lifetime of 3$\times$10$^5$\,s}
\tablenotetext{f}{From direct-mode profile fit}
\tablenotetext{g}{Carbon lifetime from this work}
\label{tab:parent}
\end{deluxetable}

The detailed photodissociative lifetimes and excess energies of the
CH$_4$ to carbon chain are not know.  \citet{huebner92} calculate the
photo rates of the various channels of CH$_4$ to its daughters and the
excess energies of these reactions.  The total solar quiet lifetime of
CH$_4$ is 1.3$\times$10$^5$\,s (7.7$\times$10$^{-6}$\,s$^{-1}$).  The
primary channel to atomic carbon is through CH$_2$.  The interstellar
and solar photo rates of CH$_2$ have been calculated by
\citet{vandishoeck96}, who find the total lifetime is
7 -- 11$\times$10$^3$\,s (9 -- 14$\times$10$^{-5}$\,s$^{-1}$).  The
majority of the CH$_2$ decays to CH, which, according to
\citet{huebner92} has a very short lifetime (108\,s; rate of
9.3$\times$10$^{-3}$\,s$^{-1}$).  Formally, a 4--generation
\citeauthor{haser57} model would best describe the \ion{C}{1} emission
from CH$_4$.  However, the progression of long to short
parent-daughter lifetimes allows us to emulate the distribution with a
standard 2-component \citeauthor{haser57} model, with the effective
parent lifetime equal to the sum of all of the lifetimes (see
\tab{parent}).  We mount a similar argument to estimate limits on the
effective ejection velocity of carbon from the chain of CH$_4$ decay.
From \citeauthor{huebner92}, the first and last links are 3 and
0.7\,\kms, respectively, which add in quadrature to find 3.1\,\kms.
Excess energies of the CH$_2$ dissociations are not quoted by
\citet{vandishoeck96}, so 3.1\,\kms\ is a lower limit to the carbon
ejection velocity from the CH$_4$ chain.  When we created the model
described below, we used 3.7\,\kms, which assumes a 2\,\kms\ ejection
velocity of CH from CH$_2$.

The CH$_3$OH lifetime is 8.7$\times$10$^4$\,s
(1.1$\times$10$^{-5}$\,s$^{-1}$) and dissociates primarily to H$_2$CO,
which lives 4600\,s \citep{huebner92}.  H$_2$CO primarily dissociates
to CO, so CH$_3$OH cannot provide the short-lived component of the
\ion{C}{1} emission.  The ejection velocities of the daughters from
the CH$_3$OH and H$_2$CO dissociations are 1.7 and 1\,\kms,
respectively, so, adding in quadrature with the carbon ejection
velocity from CO, we find a total $v_\mathrm{e}$ = 4.5\,\kms.

\begin{figure}
  \plotone{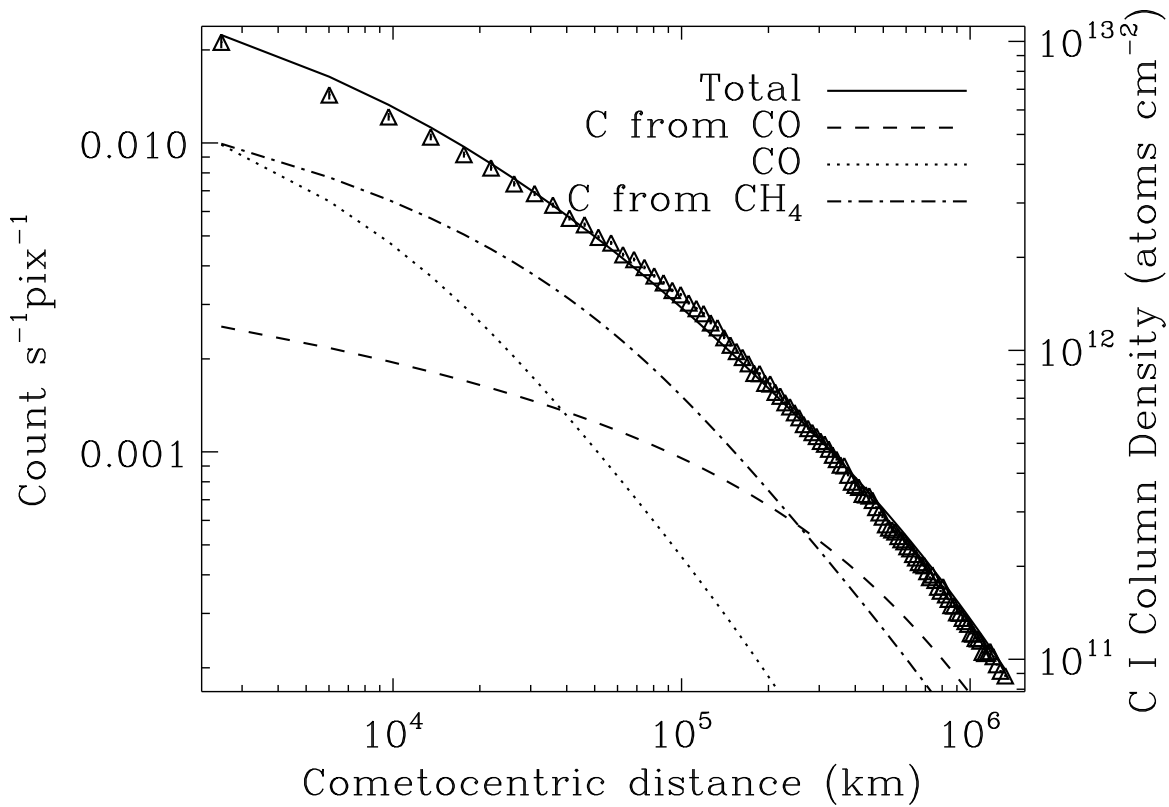}
  \caption[Comet Machholz \GALEX\ FUV radial profile]{Radial profile
  from quadrant~1 using 500 counts per bin compared to the
  three-component model discussed in the text.  Residuals of this
  model from the grism-mode data are shown in the lower right-hand
  panel of \fig{grism}.}
  \label{fig:profile}
\end{figure}

The fact that the final carbon ejection velocities we calculate for
our three candidate \ion{C}{1} parent molecules all lie within 25\% of
each other is significant.  This means that after the details of the
parent molecule dissociation are sorted out, the carbon from all of
these sources will be traveling at roughly the same speed away from
the nucleus of the comet.  At large cometocentric distances,
\citeauthor{haser57} models of these individual components will
therefore be parallel to each other.  This perhaps explains why the
radial profiles shown in \fig{profiles} and \fig{profile} follow the
characteristic \citeauthor{haser57} profile outside of \about
3$\times$10$^5$\,km despite the fact that multiple sources of
\ion{C}{1} are present.

As a final test of our analyses, we constructed a three-component
model consisting of CO Fourth Positive and two \ion{C}{1} sources, one
from CO and one from CH$_4$.  The lifetimes and ejection velocities
are discussed above and the best-fit production rates are shown in
\tab{parent}.  The grism-mode residual is shown in the lower
right-hand panel of \fig{grism}.  The model, with its constituent
parts is shown compared to the direct-mode quadrant~1 profile in
\fig{profile}.

A final adjustment to achieve the best residuals to the grism-mode
data was made by shifting the comet image 2 pixels in the positive
dispersion direction.  This corresponds to 3\arcsec, which is smaller
than the 4.5\arcsec\ \GALEX\ FUV point-spread function and reasonable
given the difficulty in finding high-precision astrometric solutions
in the dispersion direction from stellar spectra at a spectroscopic
resolving power ($\lambda/\Delta\lambda$) of 200.  

Having achieved a good residual grism-mode image, we established an
upper limit of Q(S) = 2$\times$10$^{26}$\,s$^{-1}$, assuming the
dominant parent of S is H$_2$S \citep[e.g.,][]{mcphate99} with
lifetimes from \citet{meier97}.

\section{Discussion}
\label{discussion}

We have created a model of CO and \ion{C}{1} emission that describes
the \GALEX\ direct and grism-mode observations of comet Machholz
(\figsp{grism}{profile}).  The model shows the surprising result that
carbon from CH$_4$ dominates carbon emission from CO out to \about
2$\times$10$^5$\,km.  At even greater cometocentric distances
\ion{C}{1} emission from CH$_3$OH may be significant.  Since the
carbon ejection velocities from all of these parent molecules is
likely to be comparable (\tab{parent}), the net effect of multiple
carbon parent molecules on our carbon ionization lifetime measurement
is negligible, as long as the \GALEX\ FOV samples the region of the
coma beyond the effective combined parent scale length (indicated by
the ``knee'' in the two-component \citeauthor{haser57} model).
Inspection of \fig{profile} suggests that such a knee has been reached
in the data at \about 2$\times$10$^5$\,km.

Carbon emission from CH$_3$OH, which dissociates to carbon through CO,
will everywhere have a flatter distribution than carbon emission from
CO and CH$_4$.  If this emission is significant, the flat part of the
parent distribution may flatten the total \ion{C}{1} distribution.  If
this were the case, the carbon ionization lifetime measured by our
method would be artificially long by some unknown amount.  However, we
estimate the total \ion{C}{1} emission from CH$_3$OH will be no more
than that of \ion{C}{1} from CO (\tab{parent}), so its contribution
will be smaller inside some critical radius.  This critical radius
will likely be larger than the cross-over between the CH$_4$ and CO
\ion{C}{1} sources.  If this cross-over occurs outside the \GALEX\
FOV, it will have no effect on our measurements.  If the cross-over
occurs within the \GALEX\ FOV, our estimate of the carbon ionization
lifetime could be lowered by no more than 12\% (the difference between
the outflow velocities of carbon from CO and CH$_3$OH).  This is well
within the systematic uncertainty induced by our background
subtraction (\S\ref{background}, \figp{TCI_back}, \S\ref{compare} and
\tab{rates}) and comparable to the uncertainty in our lifetime
measurement caused by our choice of $r_\mathrm{min}$ (\S\ref{haser}).

It is also interesting to note that the CO lifetime may be shorter
than 5$\times$10$^5$\,s (rate $>$ 2$\times$10$^{-6}$\,s$^{-1}$) and
that \ion{C}{1} emission from the CH$_4$ chain contributes less than
we estimate.  Using the solar maximum \citet{tozzi98} CO
photodissociation rate value and the solar wind ionization rates from
\citet{rubin09}, the CO lifetime could be as short as
3.3$\times$10$^5$\,s (3.0$\times$10$^{-6}$\,s$^{-1}$).  When we make
this change to our model we find that carbon production from CH$_4$
decreases to 1.0$\times$10$^{28}$\,s$^{-1}$ but that there is no
change in the carbon production rate from CO.  The production rate of
carbon from CO could be lowered toward the Q(C)$_\mathrm{max}$ value
quoted in \tab{parent} if we included carbon from CH$_3$OH in our
model.  As we have shown in the previous paragraphs, these adjustments
do not effect the ultimate goal of this work: the measurement of the
carbon ionization lifetime.  We look forward to continuing the
analysis of our grism data to directly measure the CO lifetime and
better constrain the production rates of the various carbon parent
molecules with these UV data.  Calculation of the excess energies of
dissociation of CH$_2$ to solar photodissociation would also be useful
in verifying our CH$_4$ chain analysis.

\section{Conclusion}
\label{conclusion}

We have used high-quality FUV observations of comet C/2004 Q2
(Machholz) collected by the \GALEX\ satellite and simple comet coma
modeling techniques to measure the lifetime of ground state atomic
carbon against ionization processes in interplanetary space.  Scaled
to 1\,AU, we measure a value of 7.1 -- 9.6$\times$10$^5$\,s
(ionization rate of 1.0 -- 1.4$\times$10$^{-6}$\,s$^{-1}$).  This
measurement compares favorably to calculations of the total ionization
lifetime from three processes (in order of importance): solar wind
proton charge exchange, solar photoionization, and solar wind electron
impact ionization, assuming the comet was in the slow solar wind
(\tab{rates}).

Comet Machholz was observed at a heliographic latitude of 30\arcdeg,
which is near the boundary between the fast and slow solar wind
\citep{bzowski03}.  Our measurement was sensitive enough to determine
that the comet was in the slow solar wind (\S\ref{discussion} and
\figp{TCI_back}).  Together with ion tail observations, observation of
the scale length of \ion{C}{1} can therefore provide diagnostics of
solar wind conditions at heliographic latitudes not often sampled by
spacecraft.

The principal measurement uncertainty in the determination of the
carbon ionization lifetime was related to the uncertainty in
determining the background in the direct-mode comet image.  The
\ion{C}{1} scale length was significantly larger than the FOV radius,
so the background could not be determined \textit{a priori} from the
comet image itself.  A background image recorded a month earlier was
used, but, as first noted by \citet{sujatha09}, variable airglow
contamination makes simple-minded background subtraction of \GALEX\
data recorded at different times highly inaccurate.  We created a
general method for the removal of airglow contamination that
accurately considers observing geometry.  This method could be
improved using more physically accurate airglow emitter profiles from
ionospheric and plasmaspheric models.

Our method of simultaneously using the \GALEX\ direct and grism-mode
observations has made it possible to measure the ionization lifetime
of carbon and begin to probe the distribution of the various carbon
parent molecules.  With careful cleaning of background sources from
the grism data, this work will continue.  In particular, it will be
possible to use our parameter optimizer to directly measure the CO
lifetime.  With the addition of optical depth effects to our
$g$-factor calculations, we will also be able to use the \GALEX\ data
to determine if CO comes from a distributed source in comet Machholz.

\GALEX\ has observed two other comets to date in the FUV: 8P/Tuttle
and C/2007 N3 (Lulin) and six comets in the NUV: C/2004 Q2 (Machholz),
9P/Tempel~1, 73P/Schwassmann-Wachmann 3 Fragments B and C, 8P/Tuttle
and C/2007 N3 (Lulin).  The \GALEX\ NUV channel includes emission from
the bright OH~3080\,\AA\ band, the CS~2576\,\AA\ band, dust continuum
and CO$^+$ features.  The combined image and grism-mode analysis
described here is also applicable to the \GALEX\ NUV data.

\acknowledgments

We thank Karl Foster for his help planning the \GALEX\ comet
observations, Tim Conrow, and the entire \GALEX\ pipeline team for
providing high quality products to work with, Jon Giorgini and the JPL
HORIZONS team for calculating high quality cometary ephemerides from
the perspective of \GALEX, and Wayne Landsman and the many
contributors to the IDLASTRO library.  \textit{GALEX} is operated for
NASA by the California Institute of Technology under NASA contract
NAS5-98034.  This work was supported as part of the \textit{GALEX}
Guest Investigator program, grant NNG05GJ93G and NNX08AE14G to the
University of Washington and NNX08AU56G to the Planetary Science
Institute and NASA Planetary Atmospheres Grant NNX09AB59G to the
University of Michigan.  This is PSI Contribution Number 493.

{\it Facilities:} \facility{GALEX}.

\bibliography{apj-jour,comets,cross_sections,Fabry_Perot,io,SHS,sun}
\bibliographystyle{apj}

\end{document}